\RequirePackage{lineno}
\documentclass[twocolumn,letterpaper,aps,prl,superscriptaddress,floatfix,nofootinbib]{revtex4-1}

\usepackage{xspace}	
\usepackage{graphicx}	
\usepackage{multirow}
\usepackage{xcolor}
\usepackage{soul}
\usepackage{ulem}

\newcommand{\beq}{\begin{equation}}
\newcommand{\eeq}{\end{equation}}

\begin{document}
\title{Evidence for Nonlinear Gluon Effects in QCD and their $A$ Dependence at STAR}
\affiliation{Abilene Christian University, Abilene, Texas   79699}
\affiliation{AGH University of Science and Technology, FPACS, Cracow 30-059, Poland}
\affiliation{Alikhanov Institute for Theoretical and Experimental Physics NRC "Kurchatov Institute", Moscow 117218}
\affiliation{Argonne National Laboratory, Argonne, Illinois 60439}
\affiliation{American University of Cairo, New Cairo 11835, New Cairo, Egypt}
\affiliation{Brookhaven National Laboratory, Upton, New York 11973}
\affiliation{University of California, Berkeley, California 94720}
\affiliation{University of California, Davis, California 95616}
\affiliation{University of California, Los Angeles, California 90095}
\affiliation{University of California, Riverside, California 92521}
\affiliation{Central China Normal University, Wuhan, Hubei 430079 }
\affiliation{University of Illinois at Chicago, Chicago, Illinois 60607}
\affiliation{Creighton University, Omaha, Nebraska 68178}
\affiliation{Czech Technical University in Prague, FNSPE, Prague 115 19, Czech Republic}
\affiliation{Technische Universit\"at Darmstadt, Darmstadt 64289, Germany}
\affiliation{ELTE E\"otv\"os Lor\'and University, Budapest, Hungary H-1117}
\affiliation{Frankfurt Institute for Advanced Studies FIAS, Frankfurt 60438, Germany}
\affiliation{Fudan University, Shanghai, 200433 }
\affiliation{University of Heidelberg, Heidelberg 69120, Germany }
\affiliation{University of Houston, Houston, Texas 77204}
\affiliation{Huzhou University, Huzhou, Zhejiang  313000}
\affiliation{Indian Institute of Science Education and Research (IISER), Berhampur 760010 , India}
\affiliation{Indian Institute of Science Education and Research (IISER) Tirupati, Tirupati 517507, India}
\affiliation{Indian Institute Technology, Patna, Bihar 801106, India}
\affiliation{Indiana University, Bloomington, Indiana 47408}
\affiliation{Institute of Modern Physics, Chinese Academy of Sciences, Lanzhou, Gansu 730000 }
\affiliation{University of Jammu, Jammu 180001, India}
\affiliation{Joint Institute for Nuclear Research, Dubna 141 980}
\affiliation{Kent State University, Kent, Ohio 44242}
\affiliation{University of Kentucky, Lexington, Kentucky 40506-0055}
\affiliation{Lawrence Berkeley National Laboratory, Berkeley, California 94720}
\affiliation{Lehigh University, Bethlehem, Pennsylvania 18015}
\affiliation{Max-Planck-Institut f\"ur Physik, Munich 80805, Germany}
\affiliation{Michigan State University, East Lansing, Michigan 48824}
\affiliation{National Research Nuclear University MEPhI, Moscow 115409}
\affiliation{National Institute of Science Education and Research, HBNI, Jatni 752050, India}
\affiliation{National Cheng Kung University, Tainan 70101 }
\affiliation{Nuclear Physics Institute of the CAS, Rez 250 68, Czech Republic}
\affiliation{Ohio State University, Columbus, Ohio 43210}
\affiliation{Institute of Nuclear Physics PAN, Cracow 31-342, Poland}
\affiliation{Panjab University, Chandigarh 160014, India}
\affiliation{Pennsylvania State University, University Park, Pennsylvania 16802}
\affiliation{NRC "Kurchatov Institute", Institute of High Energy Physics, Protvino 142281}
\affiliation{Purdue University, West Lafayette, Indiana 47907}
\affiliation{Rice University, Houston, Texas 77251}
\affiliation{Rutgers University, Piscataway, New Jersey 08854}
\affiliation{Universidade de S\~ao Paulo, S\~ao Paulo, Brazil 05314-970}
\affiliation{University of Science and Technology of China, Hefei, Anhui 230026}
\affiliation{Shandong University, Qingdao, Shandong 266237}
\affiliation{Shanghai Institute of Applied Physics, Chinese Academy of Sciences, Shanghai 201800}
\affiliation{Southern Connecticut State University, New Haven, Connecticut 06515}
\affiliation{State University of New York, Stony Brook, New York 11794}
\affiliation{Instituto de Alta Investigaci\'on, Universidad de Tarapac\'a, Arica 1000000, Chile}
\affiliation{Temple University, Philadelphia, Pennsylvania 19122}
\affiliation{Texas A\&M University, College Station, Texas 77843}
\affiliation{University of Texas, Austin, Texas 78712}
\affiliation{Tsinghua University, Beijing 100084}
\affiliation{University of Tsukuba, Tsukuba, Ibaraki 305-8571, Japan}
\affiliation{Valparaiso University, Valparaiso, Indiana 46383}
\affiliation{Variable Energy Cyclotron Centre, Kolkata 700064, India}
\affiliation{Warsaw University of Technology, Warsaw 00-661, Poland}
\affiliation{Wayne State University, Detroit, Michigan 48201}
\affiliation{Yale University, New Haven, Connecticut 06520}

\author{M.~S.~Abdallah}\affiliation{American University of Cairo, New Cairo 11835, New Cairo, Egypt}
\author{B.~E.~Aboona}\affiliation{Texas A\&M University, College Station, Texas 77843}
\author{J.~Adam}\affiliation{Brookhaven National Laboratory, Upton, New York 11973}
\author{L.~Adamczyk}\affiliation{AGH University of Science and Technology, FPACS, Cracow 30-059, Poland}
\author{J.~R.~Adams}\affiliation{Ohio State University, Columbus, Ohio 43210}
\author{J.~K.~Adkins}\affiliation{University of Kentucky, Lexington, Kentucky 40506-0055}
\author{G.~Agakishiev}\affiliation{Joint Institute for Nuclear Research, Dubna 141 980}
\author{I.~Aggarwal}\affiliation{Panjab University, Chandigarh 160014, India}
\author{M.~M.~Aggarwal}\affiliation{Panjab University, Chandigarh 160014, India}
\author{Z.~Ahammed}\affiliation{Variable Energy Cyclotron Centre, Kolkata 700064, India}
% ...

\collaboration{STAR Collaboration}\noaffiliation

\begin{abstract}

The STAR Collaboration reports measurements of back-to-back azimuthal correlations of di-$\pi^0$s produced at forward pseudorapidities ($2.6<\eta<4.0$) in $p$+$p$, $p+$Al, and $p+$Au collisions at a center-of-mass energy of 200 GeV.
We observe a clear suppression of the correlated yields of back-to-back $\pi^0$ pairs in $p+$Al and $p+$Au collisions compared to the $p$+$p$ data. The observed suppression of back-to-back pairs as a function of transverse momentum suggests nonlinear gluon dynamics arising at high parton densities. The larger suppression found in $p+$Au relative to $p+$Al collisions exhibits a dependence of the saturation scale, $Q_s^2$, on the mass number, $A$. A linear scaling of the suppression with $A^{1/3}$ is observed with a slope of $-0.09$ $\pm$ $0.01$.
\end{abstract}

	
\maketitle
The quest to understand quantum chromodynamics (QCD) processes in cold nuclear matter has in the last years revolved around the following questions:  Can we experimentally find evidence for a novel universal regime of nonlinear QCD dynamics in nuclei? What is the role of saturated strong gluon fields? And what are the degrees of freedom in this high gluon density regime?
These questions have motivated and continue to motivate theoretical efforts and experiments at facilities worldwide.

Collisions between hadronic systems, $i.e.$, $p$+$A$ and $d$+$A$ at the Relativistic Heavy Ion Collider (RHIC) provide a window to the parton distributions of nuclei at small momentum fraction $x$ (down to $10^{-3}$). Several RHIC measurements have shown that,
at forward pseudorapidities (deuteron going direction),
the hadron yields are suppressed in $d+$Au collisions relative to $p$+$p$ collisions in inclusive productions~\cite{BRAHMS:2003sns,Arsene:2004ux,Adler:2004eh,Adams:2006uz} and di-hadron correlations~\cite{Adams:2006uz,Adare:2011sc}.
The mechanisms leading to the suppression are not firmly established. The density of gluons in nucleons and nuclei increases at low $x$ due to gluon splitting. At a sufficiently small value of $x$, yet to be determined by experiments, the splitting is expected to be balanced by gluon recombination~\cite{Mueller:1985wy,Hwa:2004in}. The resulting gluon saturation~\cite{Gribov:1984tu,Armesto:2004ud,Gelis:2010nm,Albacete:2010pg,Tuchin:2009nf,Kovchegov:2012mbw,ALBACETE20141,Morreale:2021pnn} is one of the possible explanations for the suppression of forward hadron (jet) production. 
Initial- and final-state multiple scattering can determine the strength of the nuclear-induced transverse momentum imbalance for back-to-back particles~\cite{Vitev:2003xu,Vitev:2007ve,Frankfurt:2007rn,Kang:2011bp}. Energy loss in the nuclear medium is also predicted to result in a significant suppression of  forward hadron (jet) production. For $d$+$A$ the contributions from double-parton interactions to the $d$+$A$$\rightarrow \pi^0 \pi^0$X cross section are suggested as an alternative explanation for the suppression~\cite{Strikman:2010bg}. Therefore, it is important to make the same measurements in the theoretically and experimentally cleaner $p$+$A$ collisions.

Back-to-back di-hadron azimuthal angle correlations have been proposed to be one of the most sensitive probes to directly access the underlying gluon dynamics involved in hard scatterings~\cite{MARQUET200741,Zheng:2014vka}.
At a given $x$, the density of
gluons per unit transverse area is expected to be larger in
nuclei than in nucleons; thus, nuclei provide a natural
environment to study nonlinear gluon evolution~\cite{Gribov:1984tu}. Under the color glass condensate (CGC) framework~\cite{McLerran:1993ni,McLerran:1993ka,iancu2002colour}, gluons from different nucleons are predicted to amplify the total transverse gluon density by a factor of $A^{1/3}$ for a nucleus with mass number $A$.
Saturation is characterized by a transverse momentum scale, referred to as $Q_{s}$. Two modes can be identified: one weakly coupled (transverse momentum $k_{\bot} \gg Q_{s}$) and one strongly coupled ($k_{\bot} \le Q_{s}$)~\cite{Blaizot_2017}.
$Q_{s}$ of a nucleus is enhanced with respect to the nucleon at fixed values of $x$ and $Q^{2}$.
One can parametrize the gluon distributions following the Golec-Biernat W$\ddot{u}$sthoff (GBW) model \cite{GolecBiernat:1998js} with $Q^{2}_{s}\propto A^{1/3}Q^{2}_{s0}(x/x_{0})^{-\lambda}$, where $Q_{s0}$ = 1 GeV, $x_{0}$ = 3.04$\times10^{-4}$, and $\lambda$ = 0.288.
The CGC framework predicts that at forward angles (large pseudorapidities) high $x$ quarks and gluons in the nucleon interact coherently with gluons at low $x$ in the nucleus~\cite{Guzey:2004zp}.  As a result, the probability to observe the
associated hadrons is expected to be suppressed in $p(d)$+$A$ collisions compared 
to $p$+$p$, and an angular broadening of the back-to-back correlation of di-hadrons is predicted~\cite{Kharzeev:2004bw,Marquet:2007vb}. 

In this Letter, we report measurements of back-to-back azimuthal correlations of di-$\pi^0$s in $p+$Al and $p+$Au relative to $p$+$p$ collisions in the forward-pseudorapidity region ($2.6<\eta<4.0$) at $\sqrt{s_{\mathrm{_{NN}}}}$ = 200 GeV. The near-side peak mainly addresses physics related to fragmentation and is therefore not discussed in this Letter. If the suppression of correlation functions is observed in $p$+$A$ collisions, the use of different ion beams provides the opportunity to test the CGC prediction of $Q_s^2$ dependence on $A$. The data were obtained from $p$+$p$, $p+$Al, and $p+$Au collisions in 2015 with the $\pi^0$s reconstructed from photons, which were identified with the STAR forward meson spectrometer (FMS). 

The FMS is an electromagnetic calorimeter installed at the STAR experiment in the forward-pseudorapidity region~\cite{STAR:2018iyz}. It is 7 meters away from the nominal interaction point, facing the clockwise circulating RHIC proton beam, which makes the FMS response insensitive to the $p$, Al, and Au target beam remnants. The FMS is a highly-segmented octagonal shaped wall with a 40 cm $\times$ 40 cm square hole surrounding the beam pipe. It contains 1264 lead glass blocks of two different types and sizes. The 476 small cells from the inner portion each have dimensions of about 3.8 cm $\times$ 3.8 cm $\times$ 45 cm and collectively cover a pseudorapidity range from 3.3 to 4.0. The outer region surrounding the small cells is a set of 788 large cells, 5.8 cm $\times$ 5.8 cm $\times$ 60 cm in size, covering a pseudorapidity range from 2.6 to 3.3. 

\begin{figure}[!ht]
\vspace{-0.0cm}
\hspace{-0.5cm}
  \includegraphics[width=0.95\linewidth]{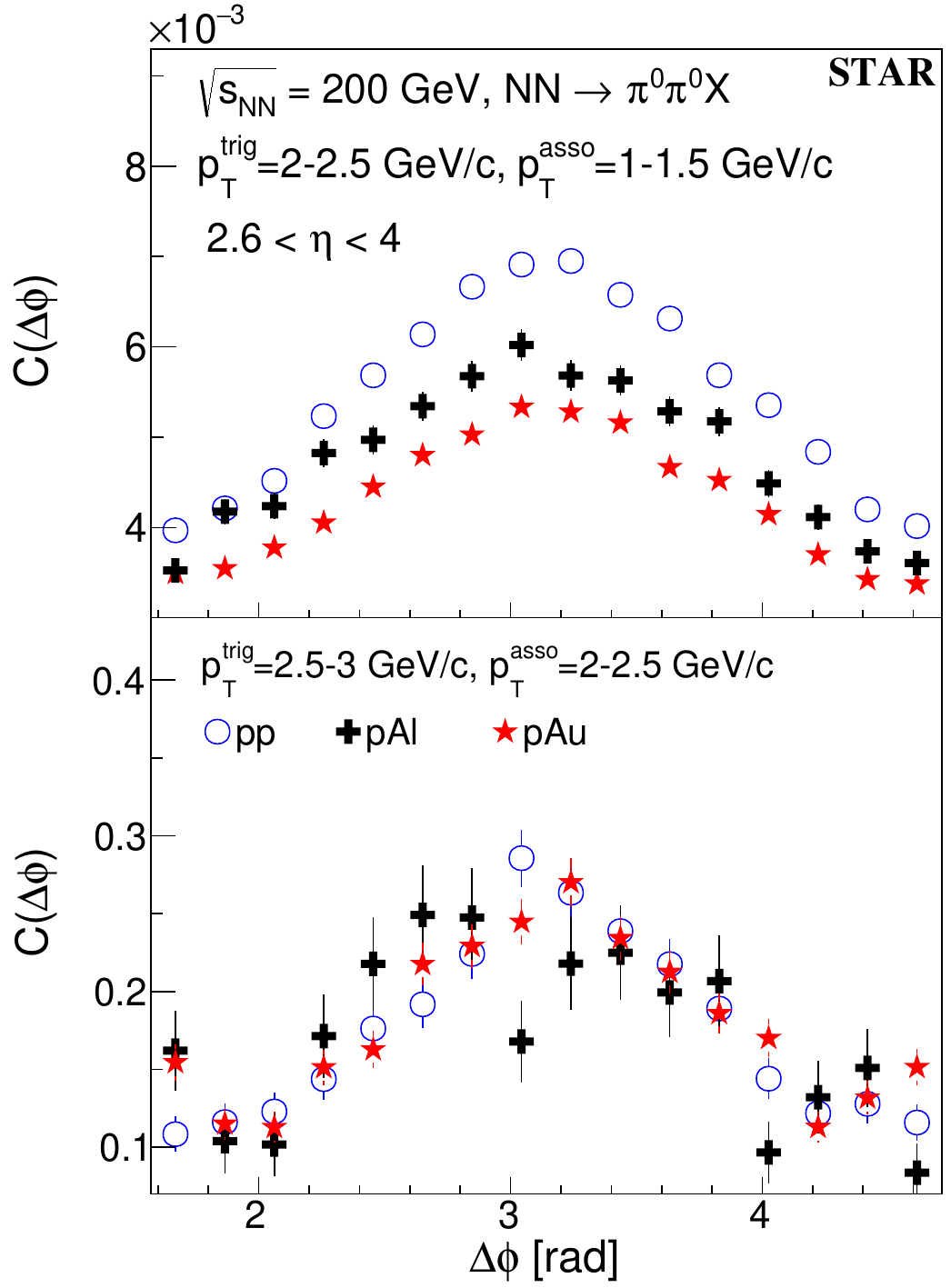}
\caption{(color online). 
Comparison of the correlation functions (corrected for nonuniform detector efficiency in $\phi$; not corrected for the absolute detection efficiency) vs. azimuthal angle difference between forward ($2.6<\eta<4.0$) $\pi^{0}$s in $p$+$p$, $p+$Al, and $p+$Au collisions at $\sqrt{s_{\mathrm{_{NN}}}}=200$ GeV.
Upper panel: the trigger $\pi^0$'s $p_{T}$ ($p^{\mathrm{trig}}_T$) = 2$-$2.5 GeV/$c$ and the associated $\pi^0$'s $p_{T}$ ($p^{\mathrm{asso}}_T$) = 1$-$1.5 GeV/$c$; according to the fit described in the text, the area$\times10^{3}$ (width) of the correlation in $p$+$p$, $p$+Al, and $p$+Au collisions are $5.67 \pm 0.12$ ($0.68 \pm 0.01$), $4.15 \pm 0.24$ ($0.68 \pm 0.03$), and $3.30 \pm 0.07$ ($0.64 \pm 0.01$), respectively.
Bottom panel: $p^{\mathrm{trig}}_T$ = 2.5$-$3 GeV/$c$ and $p^{\mathrm{asso}}_T$ = 2$-$2.5 GeV/$c$; the area$\times10^{3}$ (width) of the correlation in $p$+$p$, $p$+Al, and $p$+Au collisions are $0.18 \pm 0.01$ ($0.47 \pm 0.03$), $0.13\pm 0.03$ ($0.51 \pm 0.07$), and $0.15 \pm 0.01$ ($0.45 \pm 0.03$), respectively.
}
  \label{fig:A_corr}
\end{figure}

The collision events are triggered by the FMS itself, based on the transverse energy. 
The FMS board sum triggers~\cite{STAR:2018iyz}, which demand that the energy sum in localized overlapping areas is above particular thresholds, are used in the analysis.
To remove the beam background, the multiplicity at the Time of Flight  detector ($|\eta|<0.9$)~\cite{STAR:2002eio} is required to be above 2 and the number of tiles firing at the backward (aluminum and gold going direction) beam beam counter~\cite{Whitten:2008zz} (BBC, $-5.0<\eta<-3.3$) is above 0.
The energy and transverse momentum, $p_{T}$, of the photon candidates are required to be above 1 GeV and 0.1 GeV/$c$, respectively. 
The energy asymmetry of $\pi^0$'s photon components $|\frac{E_{1}-E_{2}}{E_{1}+E_{2}}|$ is required to be under 0.7 to reduce the combinatoric background which peaks near 1; this selection is commonly utilized in reconstructing $\pi^0$s with the FMS~\cite{STAR:2020grs,STAR:2020nnl}. The selected invariant mass range of the $\pi^0$ candidates is between 0.07 and 0.2 GeV/$c^{2}$. 

The correlation function, $C(\Delta\phi)$, is defined as $C(\Delta\phi) = \frac{N_\mathrm{pair}(\Delta\phi)}{N_\mathrm{trig}\times \Delta \phi _\mathrm{bin}}$, where $N_\mathrm{pair}$ is the yield of the correlated
trigger and associated $\pi^0$ pairs, $N_{\mathrm{trig}}$ is the
trigger $\pi^0$ yield, $\Delta\phi$ is the azimuthal angle difference between the trigger $\pi^0$ and associated $\pi^0$,
and $\Delta\phi_{\mathrm{bin}}$ is the bin width of $\Delta\phi$ distribution.
In each pair, the trigger $\pi^0$ is the one with the higher $p_{T}$ value, $p_{T}^{\mathrm{trig}}$, and the associated $\pi^0$ is the one with the lower $p_{T}$ value, $p_{T}^{\mathrm{asso}}$. To remove the correlation induced by asymmetric detector effects, the measured correlation functions shown in this Letter are corrected through dividing them by the correlation functions computed for mixed events. 
$\Delta\phi$ distributions of two $\pi^0$s produced in different events are extracted from the $\phi$ distributions of the trigger $\pi^0$s and the associated $\pi^0$s. The correlation for mixed events is the $\Delta\phi$ distribution normalized by $N_{\mathrm{bin}}$/$N^{\mathrm{\mathrm{mix}}}_{\mathrm{pair}}$, where $N_{\mathrm{bin}}$ is the number of bins in $\Delta\phi$ and $N^{\mathrm{mix}}_{\mathrm{pair}}$ is the number of $\pi^0$ pairs for mixed events. The correlations are not corrected for the absolute detection efficiency. The corrected correlation function is fitted from $\Delta\phi = -\pi/2$ to $\Delta\phi = 3\pi/2$ with two individual Gaussians at the near- ($\Delta\phi=$ 0) and away-side ($\Delta\phi=\pi$) peak, together with a constant for the pedestal. The area of the away-side peak is the integral of the correlation function from $\Delta\phi = \pi/2$ to $\Delta\phi = 3\pi/2$ after pedestal subtraction, describing the back-to-back $\pi^0$ yields per trigger particle; the corresponding width is defined as the $\sigma$ of the away-side peak according to the fit.

\begin{figure}[!ht]
\vspace{-0.0cm}
\hspace{-0.5cm}
  \includegraphics[width=0.95\linewidth]{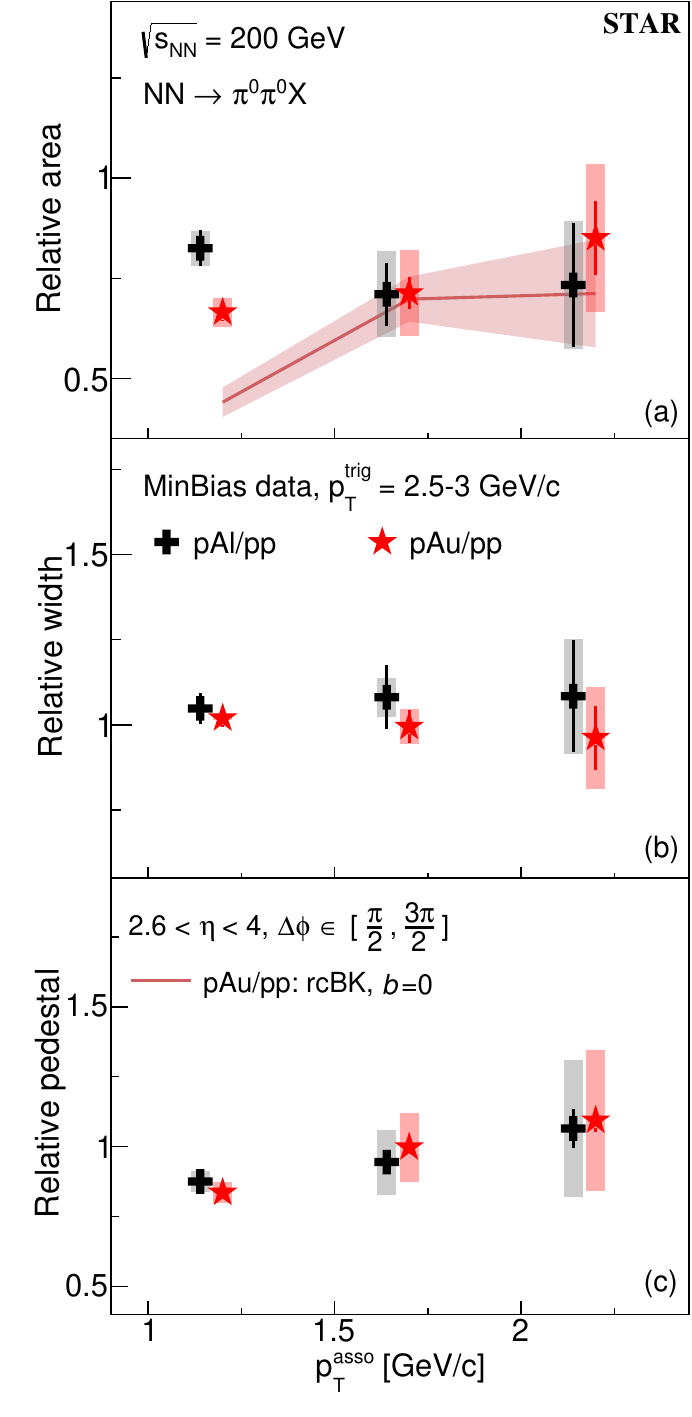}
  \caption{(color online).
Relative area (a), relative width (b), and relative pedestal (c) of back-to-back di-$\pi^0$ correlations at forward pseudorapidities ($2.6<\eta<4.0$) in $p+$Al and $p+$Au with respect to $p$+$p$ collisions for $p_{T}^{\mathrm{trig}}$ = 2.5$-$3 GeV/$c$ as a function of $p_{T}^{\mathrm{asso}}$.
The vertical bars indicate the statistical uncertainties and the vertical bands indicate the point-to-point systematic uncertainties. The horizontal width of the bands is chosen for visual clarity and does not reflect the uncertainty. The points of $p+$Al collisions are slightly offset in $p_{T}^{\mathrm{asso}}$ for visual clarity.
The theory prediction based on the rcBK model \protect\cite{Albacete:2018ruq} is calculated for an impact parameter $b = 0$.}
  \label{fig:sum_pppA}
\end{figure}

Figure~\ref{fig:A_corr} shows 
the comparison of $C(\Delta\phi)$
for forward back-to-back $\pi^0$ pairs
observed in $p$+$p$, $p+$Al, and $p+$Au collisions at $\sqrt{s_{\mathrm{_{NN}}}}=200$ GeV.
In the upper panel, in the low-$p_T$ regime, a clear suppression is observed in $p$+$A$ compared to the $p$+$p$ data. The back-to-back $\pi^0$ yields per trigger in $p+$Au ($p+$Al) are suppressed by about a factor of 1.7 (1.4) with respect to $p$+$p$ collisions.  
Larger suppression in $p+$Au relative to $p+$Al at the same collision energy supports an $A$ dependence of $Q_s^2$ as predicted in references \cite{McLerran:1993ni,Kharzeev:2004bw}. The suppression decreases with increasing $p_T$ of the $\pi^0$s. From the bottom panel of Fig.~\ref{fig:A_corr}, the suppression is found to be weaker compared to the low-$p_{T}$ range in $p+$Au collisions. 
The area, width, and pedestal in $p$+$p$, $p+$Al, and $p+$Au collisions with full di-$\pi^0$ $p_{T}$ combinations can be found in the supplemental material~\cite{Supple}.

The parton momentum fraction $x$ with respect to the nucleon inside the nucleus is proportional to the $p_{T}$ of the two $\pi^0$s; $Q$ can be approximated as the average $p_{T}$ of the two $\pi^0$s. Varying the gluon density in $x$ and $Q^{2}$ can be achieved by changing the $p_{T}$ of the two $\pi^0$s at forward pseudorapidities.
The low $x$ and $Q^{2}$ regime where the gluon density is large and expected to be saturated, can be accessed by probing low-$p_{T}$ $\pi^0$s; when $p_{T}$ is high, $x$ ($Q^{2}$) is not sufficiently small to reach the nonlinear regime. The simulated $x$ and $Q^{2}$ distributions in $p$+$p$ collisions can be found in the supplemental material~\cite{Supple}. For the lowest $p_{T}$ bin that can be measured with the FMS, $p_{T}^{\mathrm{trig}}$ = 1.5$-$2 GeV/$c$ and $p^{\mathrm{asso}}_T$ = 1$-$1.5 GeV/$c$, the probed $x_{2}$ covers a wide range from $10^{-4}$ to $\sim0.5$. The mean values of $x_{2}$ and $Q^{2}$ for this bin are 0.05 and 2.2 GeV$^{2}$, respectively.
For the highest $p_{T}$ bin, $p_{T}^{\mathrm{trig}}$ = 3$-$5 GeV/$c$ and $p^{\mathrm{asso}}_{T}$ = 2$-$2.5 GeV/$c$, the mean value of $x_{2}$ is 0.1 and $Q^{2}$ is 4.6 GeV$^{2}$.

In Fig.~\ref{fig:sum_pppA},
the area, width, and pedestal ratios of back-to-back di-$\pi^0$ correlations in $p+$Al and $p+$Au relative to $p$+$p$ collisions are shown as a function of $p^{\mathrm{asso}}_T$.
The systematic uncertainties of the area, width, and pedestal are estimated from nonuniform detector efficiency for each collision system as a function of $\phi$. 
A data driven Monte Carlo method was performed bin by bin in $p_{T}$ to determine the systematic uncertainties of the area, width, and the pedestal. 
An input correlation, without detector effects, was sampled by two Gaussians at the near-/away-side peaks and a constant for pedestal.
A correlation with detector effects included was obtained by weighting the $\phi$ distributions with the data and then a mixed-event correction was applied to the correlation.
The difference between the input and the corrected correlations defines the estimated systematic uncertainties, which serves as a closure test.
The systematic uncertainty depends on $p_{T}$ and rarely depends on collision system.
The systematic uncertainties of the relative area obtained at $p^{\mathrm{asso}}_T$ = 1$-$1.5 GeV/$c$, 1.5$-$2 GeV/$c$, and 2$-$2.5 GeV/$c$ are around 5\%, 15\%, and 22\%, respectively, for $p^{\mathrm{trig}}_T$ = 2.5$-$3 GeV/$c$.
The corresponding systematic uncertainties of the relative width are 0.1\%, 5\%, and 16\%. 
The corresponding systematic uncertainties of the relative pedestal are 4\%, 12\%, and 23\%.

Theoretical calculations\footnote{Calculations are not presented since Refs.~\cite{Stasto:2011ru,Lappi:2012nh} are for different collision systems and Ref.~\cite{Stasto:2018rci} applies a different normalization method.} from Ref.~\cite{Albacete:2018ruq} predict the area ratio of central $p+$Au collisions (impact parameter $b= 0$) relative to $p$+$p$ collisions and are shown in Fig.~\ref{fig:sum_pppA}(a). In this model, the gluon content of the saturated nuclear target is described with transverse-momentum-dependent (TMD) gluon distributions and the small-$x$ evolution is calculated numerically by solving the nonlinear Balitsky-Kovchegov equation~\cite{Balitsky:1995ub,Kovchegov:1999yj} with running coupling corrections (rcBK).
No predictions of width or pedestal are shown, since the model currently does not take into account soft gluon radiation as well as several other factors that affect the width, and it does not provide predictions of pedestal.
At low $p^{\mathrm{asso}}_T$, the rcBK model predicts a larger suppression in central collisions than peripheral collisions. This 
explains the deviation between the MinBias $p+$Au data 
and the predictions at $b=0$. We will present a detailed  
study of the centrality dependence in a separate paper following this Letter.

In Fig.~\ref{fig:sum_pppA}(b), the Gaussian widths of the di-$\pi^0$ correlation peaks remain the same between $p$+$p$ and $p+$A collisions for the different $p^{\mathrm{asso}}_{T}$ ranges, i.e., the broadening predicted in the CGC framework in Refs.~\cite{Kharzeev:2004bw,Marquet:2007vb} is not observed. This observation is in agreement with a similar measurement in $d+$Au collisions by the PHENIX experiment~\cite{Adare:2011sc} and $p+$Pb collisions by the ATLAS experiment~\cite{ATLAS:2019jgo}. In Fig.~\ref{fig:sum_pppA}(c), the pedestal in $p+$A is slightly lower than in $p$+$p$ collisions at low $p^{\mathrm{asso}}_{T}$. At high $p^{\mathrm{asso}}_{T}$, the pedestals from $p$+$p$ and $p+$A collisions are virtually identical. Note that the measured pedestal 
in $d+$Au is 2$-$3 times higher than in $p$+$p$ collisions~\cite{Adare:2011sc}.  This observation can provide insight in the contribution of multiple parton interactions to di-hadron correlation in $d+$Au collisions~\cite{Strikman:2010bg,Lappi:2012nh}.

\begin{figure}[h]
  \includegraphics[width=0.95\linewidth]{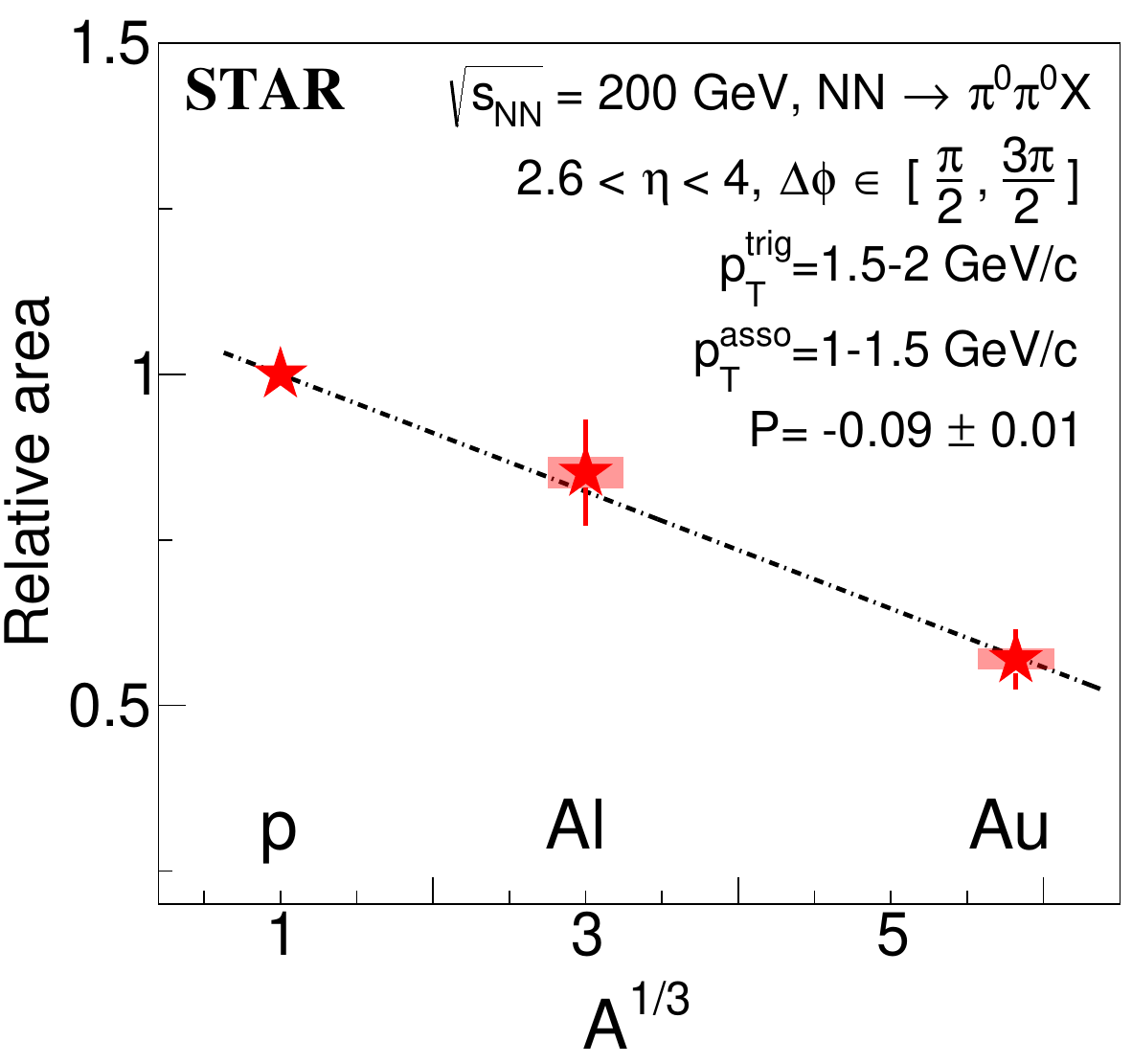}
  \caption{(color online).
  Relative area of back-to-back di-$\pi^0$ correlations at forward pseudorapidities ($2.6<\eta<4.0$) in $p+$Au and $p+$Al respect to $p$+$p$ collisions for $p^{\mathrm{trig}}_{T}$ = 1.5$-$2 GeV/$c$ and $p^{\mathrm{asso}}_{T}$ = 1$-$1.5 GeV/$c$. The vertical bars for the Al and Au ratios indicate the statistical uncertainties and the vertical bands indicate the point-to-point systematic uncertainties.
The horizontal width of the bands is chosen for visual clarity and does not reflect the uncertainty. The data points are fitted by a linear function, whose slope ($P$) is found to be $-0.09$ $\pm$ $0.01$.
}
  \label{fig:Adep}
\end{figure}

The STAR experiment performed a unique measurement of the $A$-dependence in back-to-back di-$\pi^0$ correlations at forward pseudorapidities. The relative area in $p+$Au and $p+$Al with respect to $p$+$p$ collisions is shown in Fig.~\ref{fig:Adep} as a function of $A^{1/3}$; the systematic uncertainty is around 3\% at $p^{\mathrm{trig}}_{T}$ = 1.5$-$2 GeV/$c$ and $p^{\mathrm{asso}}_{T}$ = 1$-$1.5 GeV/$c$. Nonlinear effects are found largest in the lowest $p_{T}$ range and no strong $p_{T}^{\mathrm{trig}}$ dependence is observed.
The ratio for $A = 1$ has no uncertainty since the numerator and denominator are fully correlated.
A specific $p_{T}$ range probes the suppression in $p+$Au and $p+$Al collisions in the same $x$-$Q^{2}$ phase space. Therefore, the suppression is dominantly influenced by $A$ according to the GBW model \cite{GolecBiernat:1998js}.
A linear dependence of the suppression as a function of $A^{1/3}$ is observed within the uncertainties in Fig.~\ref{fig:Adep}, the slope ($P$) is found to be $-0.09$ $\pm$ $0.01$.

In summary, the measurements of azimuthal correlations of di-$\pi^0$s at forward pseudorapidities are performed using 2015 STAR 200 GeV $p$+$p$, $p+$Al, and $p+$Au data. 
Results of the back-to-back correlations are given as a function of $p_{T}$, with the trigger $\pi^{0}$ in the range of 1.5 $< p^{\mathrm{trig}}_T <$ 5 GeV/$c$ and the associated $\pi^0$ in the range of 1 $< p^{\mathrm{asso}}_T <$ 2.5 GeV/$c$.  A clear suppression of back-to-back yields is observed in $p$+$A$ compared to $p$+$p$ data for pairs probing small $x$ (and $Q^{2}$) with low $p_{T}$. 
The present results are the first measurements of the $A$-dependence of this nuclear effect; the suppression is enhanced with higher $A$ and scales with $A^{1/3}$. 
No increase in the width of the azimuthal angular correlation is seen within experimental uncertainties. 
The stable pedestal in $p$+$A$ and $p$+$p$ collisions provides opportunities to understand the contributions from multiple parton scatterings in $d$+$A$ collisions.

We thank the RHIC Operations Group and RCF at BNL, the NERSC Center at LBNL, and the Open Science Grid consortium for providing resources and support.  This work was supported in part by the Office of Nuclear Physics within the U.S. DOE Office of Science, the U.S. National Science Foundation, National Natural Science Foundation of China, Chinese Academy of Science, the Ministry of Science and Technology of China and the Chinese Ministry of Education, the Higher Education Sprout Project by Ministry of Education at NCKU, the National Research Foundation of Korea, Czech Science Foundation and Ministry of Education, Youth and Sports of the Czech Republic, Hungarian National Research, Development and Innovation Office, New National Excellency Programme of the Hungarian Ministry of Human Capacities, Department of Atomic Energy and Department of Science and Technology of the Government of India, the National Science Centre of Poland, the Ministry  of Science, Education and Sports of the Republic of Croatia, German Bundesministerium f\"ur Bildung, Wissenschaft, Forschung and Technologie (BMBF), Helmholtz Association, Ministry of Education, Culture, Sports, Science, and Technology (MEXT) and Japan Society for the Promotion of Science (JSPS).

\bibliography{ref}

\newpage
\setcounter{figure}{0}
\onecolumngrid
\section{Supplemental material}

\begin{table*}[!ht]
\centering 
\setlength\tabcolsep{3pt}
\addtolength{\tabcolsep}{3pt}
\begin{tabular}{c c  c  c c c} \hline
$p^{\mathrm{asso}}_{T}$ [GeV/$c$]&&&$p^{\mathrm{trig}}_{T}$ [GeV/$c$] &&Parameter\\
\multirow{6}{4em}{1.0-1.5}&1.5-2.0 &2.0-2.5  & 2.5-3.0 &3.0-5.0 & \\
\hline
& 4.51$\pm$0.15 &5.67$\pm$0.12  & 5.34$\pm$0.11 &5.58$\pm$0.11& Area$\times10^{3}$\\ 
&0.59$\pm$0.02 & 0.68$\pm$0.01 & 0.64$\pm$0.01 & 0.67$\pm$0.01 &Width\\
&29.47$\pm$0.26 & 23.10$\pm$0.17 & 23.14$\pm$0.17 & 21.68$\pm$0.17& Pedestal$\times10^{3}$\\
\hline
\multirow{3.4}{4em}{1.5-2.0}&  &1.03$\pm$0.03  & 0.91$\pm$0.03 &1.09$\pm$0.04& Area$\times10^{3}$\\ 
& & 0.61$\pm$0.02 & 0.55$\pm$0.02 & 0.66$\pm$0.02 &Width\\
& & 3.15$\pm$0.05 & 3.10$\pm$0.05 & 2.51$\pm$0.06& Pedestal$\times10^{3}$\\
\hline
\multirow{3.4}{4em}{2.0-2.5}&  & & 0.18$\pm$0.01 &0.23$\pm$0.02& Area$\times10^{3}$\\ 
& & & 0.47$\pm$0.03 & 0.61$\pm$0.04 &Width\\
& & & 0.74$\pm$0.02 & 0.50$\pm$0.02 & Pedestal$\times10^{3}$\\
\hline
\end{tabular} 
\caption{The area, width, and pedestal according to the fit of forward ($2.6<\eta<4.0$) back-to-back di-$\pi^0$ correlations in $p+$$p$ collisions for various $p_{T}$ ranges. \label{tab:parameter_ppMB} }
\end{table*}

\begin{table*}[!ht]
\centering 
\setlength\tabcolsep{3pt}
\addtolength{\tabcolsep}{3pt}
\begin{tabular}{c c  c  c c c} \hline
$p^{\mathrm{asso}}_{T}$ [GeV/$c$]&&&$p^{\mathrm{trig}}_{T}$ [GeV/$c$] &&Parameter\\
\multirow{6}{4em}{1.0-1.5}&1.5-2.0 &2.0-2.5  & 2.5-3.0 &3.0-5.0 & \\
\hline
& 3.84$\pm$0.28 &4.15$\pm$0.24  & 4.40$\pm$0.22 &4.07$\pm$0.22& Area$\times10^{3}$\\ 
&0.60$\pm$0.04 & 0.68$\pm$0.03 & 0.67$\pm$0.03 & 0.70$\pm$0.03 &Width\\
&27.30$\pm$0.48 & 21.84$\pm$0.36 & 20.25$\pm$0.34 & 19.05$\pm$0.33& Pedestal$\times10^{3}$\\
\hline
\multirow{3.4}{4em}{1.5-2.0}&  &0.69$\pm$0.07  & 0.65$\pm$0.07 &0.69$\pm$0.06& Area$\times10^{3}$\\ 
& & 0.53$\pm$0.04 & 0.60$\pm$0.05 & 0.59$\pm$0.04 &Width\\
& & 3.77$\pm$0.01 & 2.92$\pm$0.11 & 2.65$\pm$0.09& Pedestal$\times10^{3}$\\
\hline
\multirow{3.4}{4em}{2.0-2.5}&  & & 0.13$\pm$0.03 &0.29$\pm$0.09& Area$\times10^{3}$\\ 
& & & 0.51$\pm$0.07 & 0.96$\pm$0.19 &Width\\
& & & 0.79$\pm$0.05 & 0.42$\pm$0.11 & Pedestal$\times10^{3}$\\
\hline
\end{tabular} 
\caption{The area, width, and pedestal according to the fit of forward ($2.6<\eta<4.0$) back-to-back di-$\pi^0$ correlations in $p+$Al collisions for various $p_{T}$ ranges. \label{tab:parameter_pAlMB} }
\end{table*}

\begin{table*}[!ht]
\centering 
\setlength\tabcolsep{3pt}
\addtolength{\tabcolsep}{3pt}
\begin{tabular}{c c  c  c c c} \hline
$p^{\mathrm{asso}}_{T}$ [GeV/$c$]&&&$p^{\mathrm{trig}}_{T}$ [GeV/$c$] &&Parameter\\
\multirow{6}{4em}{1.0-1.5}&1.5-2.0 &2.0-2.5  & 2.5-3.0 &3.0-5.0 & \\
\hline
& 2.57$\pm$0.08 &3.30$\pm$0.07  & 3.55$\pm$0.09 &3.17$\pm$0.09& Area$\times10^{3}$\\ 
&0.49$\pm$0.01 & 0.64$\pm$0.01 & 0.65$\pm$0.01 & 0.64$\pm$0.01 &Width\\
&25.68$\pm$0.15 & 20.46$\pm$0.11 & 19.33$\pm$0.14 & 18.54$\pm$0.13& Pedestal$\times10^{3}$\\
\hline
\multirow{3.4}{4em}{1.5-2.0}&  &0.76$\pm$0.03  & 0.65$\pm$0.03 &0.75$\pm$0.04& Area$\times10^{3}$\\ 
& & 0.62$\pm$0.02 & 0.55$\pm$0.02 & 0.71$\pm$0.03 &Width\\
& & 3.26$\pm$0.04 & 3.09$\pm$0.05 & 2.56$\pm$0.06& Pedestal$\times10^{3}$\\
\hline
\multirow{3.4}{4em}{2.0-2.5}&  & & 0.15$\pm$0.01 &0.18$\pm$0.02& Area$\times10^{3}$\\ 
& & & 0.45$\pm$0.03 & 0.71$\pm$0.07 &Width\\
& & & 0.81$\pm$0.02 & 0.63$\pm$0.03 & Pedestal$\times10^{3}$\\
\hline
\end{tabular} 
\caption{The area, width, and pedestal according to the fit of forward ($2.6<\eta<4.0$) back-to-back di-$\pi^0$ correlations in $p+$Au collisions for various $p_{T}$ ranges. \label{tab:parameter_pAuMB} }
\end{table*}

\newpage

\begin{figure}[!ht]
  \includegraphics[width=1.0\linewidth]{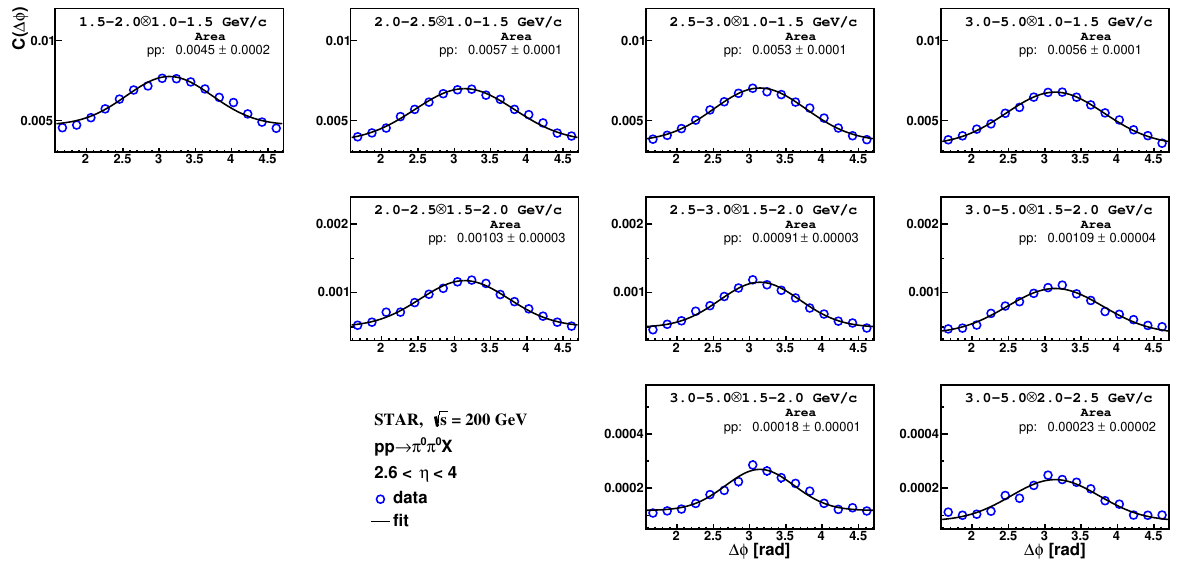}
\caption{(color online).
The forward ($2.6<\eta<4.0$) back-to-back di-$\pi^0$ correlations in  $p+$$p$ collisions for various $p_{T}$ ranges. 
The value for the area of the peak obtained from the Gaussian fit is given in each panel.
}
  \label{fig:Cpp}
\end{figure}

\begin{figure}[!ht]
  \includegraphics[width=1.0\linewidth]{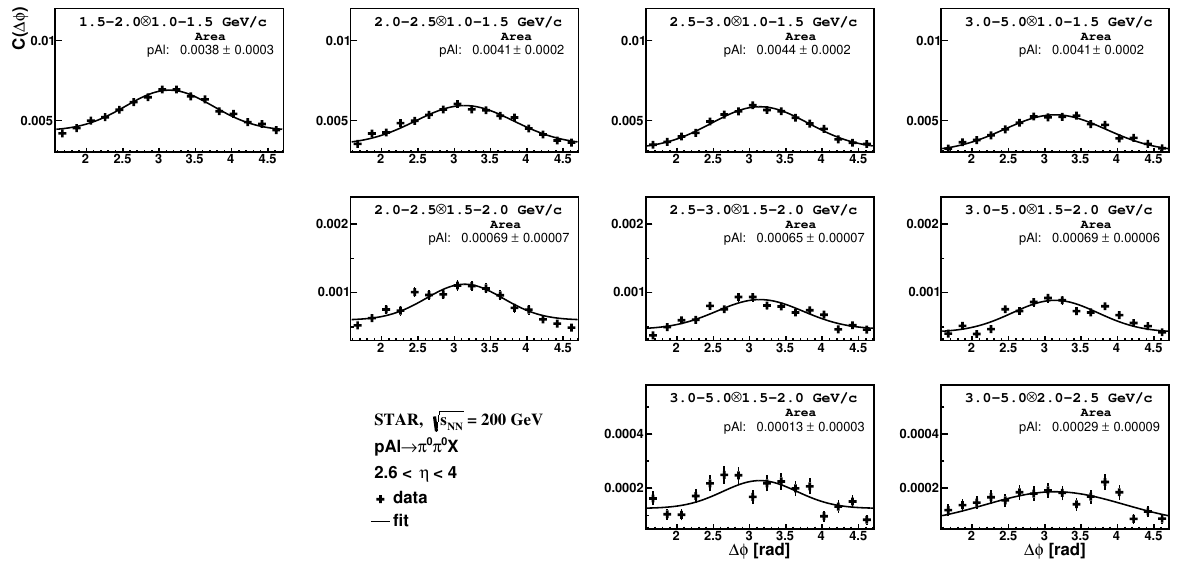}
\caption{(color online).
The forward ($2.6<\eta<4.0$) back-to-back di-$\pi^0$ correlations in  $p+$Al collisions for various $p_{T}$ ranges. 
The value for the area of the peak obtained from the Gaussian fit is given in each panel.
}
  \label{fig:Cpal}
\end{figure}

\begin{figure}[!ht]
  \includegraphics[width=1.0\linewidth]{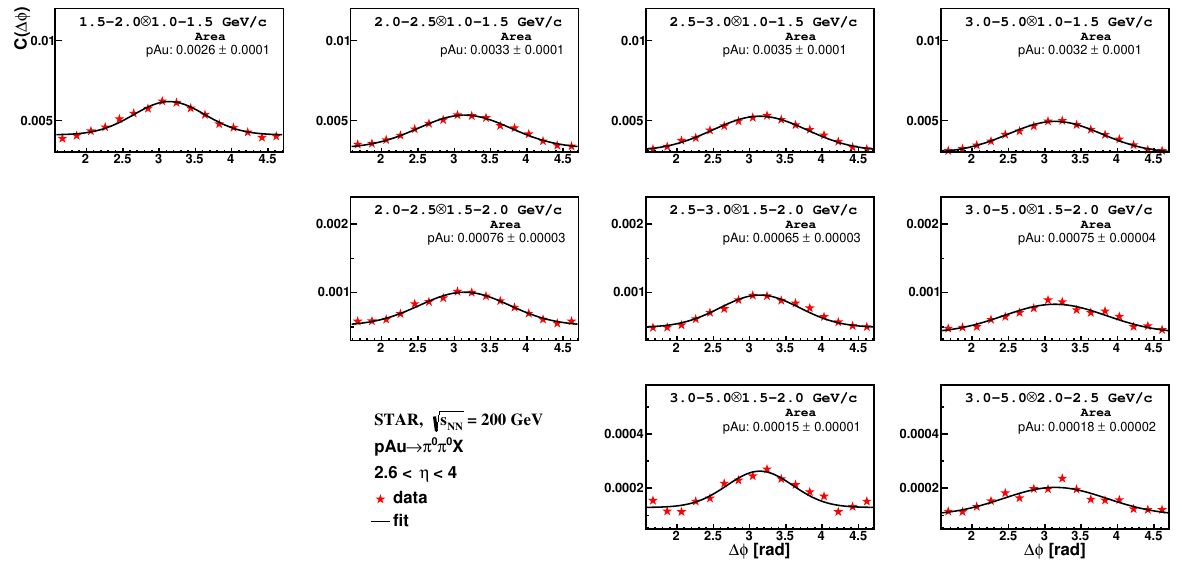}
\caption{(color online).
The forward ($2.6<\eta<4.0$) back-to-back di-$\pi^0$ correlations in  $p+$Au collisions for various $p_{T}$ ranges. 
The value for the area of the peak obtained from the Gaussian fit is given in each panel.
}
  \label{fig:Cpau}
\end{figure}

\newpage
\begin{figure}[!ht]
  \includegraphics[width=1.0\linewidth]{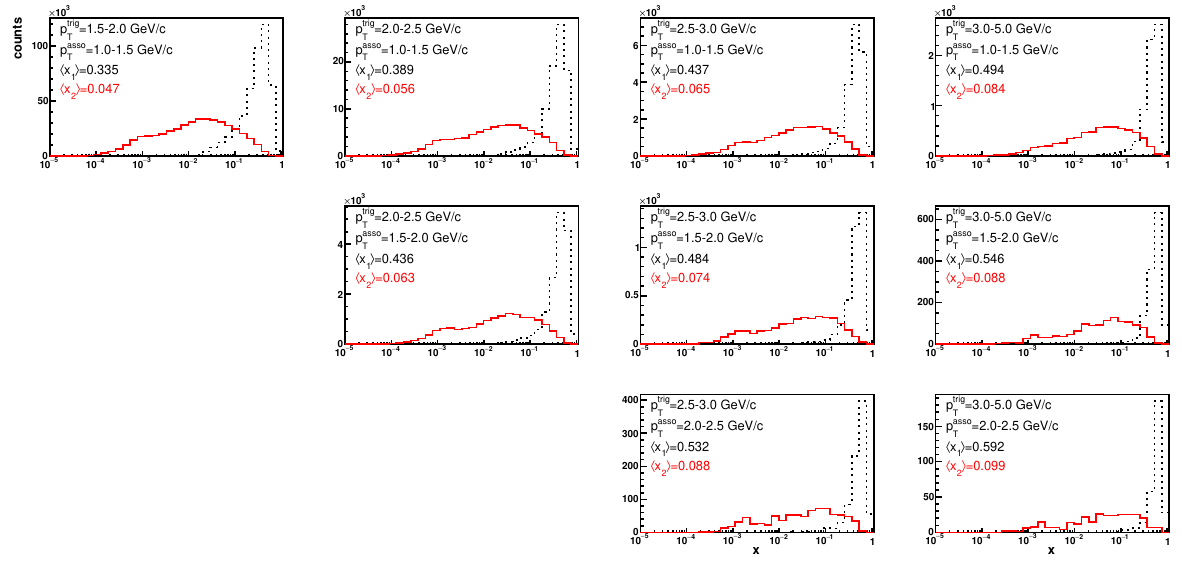}
\caption{(color online).
Monte Carlo simulations of $x_{1}$ and $x_{2}$ distributions in various $p_{T}$ ranges for $pp\rightarrow\pi^{0}\pi^{0}$X collisions at $\sqrt{s}$ = 200 GeV. The pseudorapidity range of the outgoing $\pi^{0}$s is from 2.6 to 4. The mean values of $x_{1}$ and $x_{2}$ are given in each panel.
}
  \label{fig:Area}
\end{figure}

\begin{figure}[!ht]
  \includegraphics[width=1.0\linewidth]{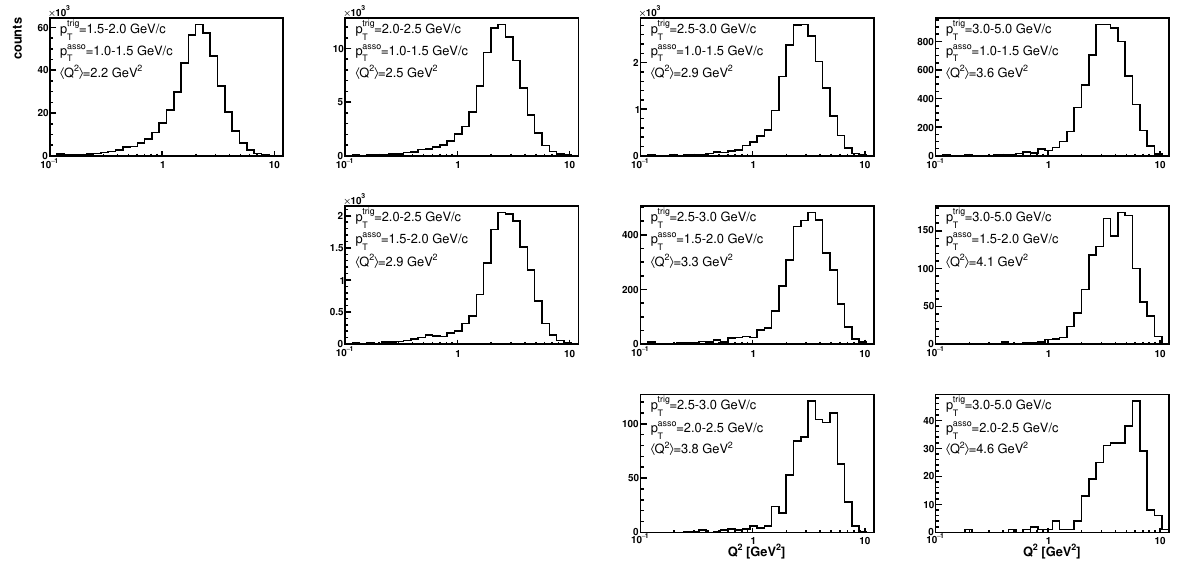}
\caption{
Monte Carlo simulations of $Q^{2}$ in various $p_{T}$ ranges for $pp\rightarrow\pi^{0}\pi^{0}$X collisions at $\sqrt{s}$ = 200 GeV. The pseudorapidity range of the outgoing $\pi^{0}$s is from 2.6 to 4. The mean value of $Q^{2}$ is given in each panel.
}
  \label{fig:Area}
\end{figure}

\end{document}